\documentclass[12pt]{article}
\usepackage{a4wide}
\usepackage{latexsym}
\usepackage{axodraw}

\usepackage{pslatex}
\usepackage[latin1]{inputenc}
\usepackage[T1]{fontenc}

\def\bq{\begin{eqnarray}}
\def\eq{\end{eqnarray}}

\def\eps{\varepsilon}
\def\v{\verb}

\newlength{\dinwidth} \newlength{\dinmargin}
\begin{document}
\thispagestyle{empty}

\begin{flushright}
  UPRF-2002-01
\end{flushright}

\vspace{1.5cm}

\begin{center}
  {\Large\bf Symbolic Expansion of Transcendental Functions\\}
  \vspace{1cm}
  {\large Stefan Weinzierl}\\
  \vspace{1cm}
  {\small {\em Dipartimento di Fisica, Universit\`a di Parma,\\
       INFN Gruppo Collegato di Parma, 43100 Parma, Italy}} \\
\end{center}

\vspace{2cm}

\begin{abstract}\noindent
  {%
    Higher transcendental function occur frequently in the
    calculation of Feynman integrals in quantum field theory.
    Their expansion in a small parameter is a non-trivial task.
    We report on a computer program which allows the systematic
    expansion of certain classes of functions.
    The algorithms are based on the Hopf algebra of nested sums.
    The program is written in C++ and uses the
    GiNaC library.
   }
\end{abstract}

\vspace*{\fill}

\newpage 

{\bf\large PROGRAM SUMMARY}
\vspace{4mm}
\begin{sloppypar}
\noindent   {\em Title of program\/}: nestedsums \\[2mm]
   {\em Version\/}: 1.0 \\[2mm]
   {\em Catalogue number\/}: \\[2mm]
   {\em Program obtained from\/}: {\tt http://www.fis.unipr.it/\~{}stefanw/nestedsums} \\[2mm]
   {\em E-mail\/}: {\tt stefanw@fis.unipr.it} \\[2mm]
   {\em License\/}: GNU Public License \\[2mm]
   {\em Computers\/}: all \\[2mm]
   {\em Operating system\/}: all \\[2mm]
   {\em Program language\/}: {\tt C++     } \\[2mm]
   {\em Memory required to execute\/}: 
         depending on the complexity of the problem, 
         at least 64 MB RAM recommended   \\[2mm]
   {\em Other programs called\/}: GiNaC 0.8.3, a library for symbolic calculations in C++, 
         is required.
         It is available from  {\tt http://www.ginac.de}.\\
         Optionally, the program can also be used interactively. 
         In this case the program gTybalt, available from 
         {\tt http://www.fis.unipr.it/\~{}stefanw/gtybalt.html} is needed. \\[2mm]
   {\em External files needed\/}: none \\[2mm]
   {\em Keywords\/}:  Multiple polylogarithms, Feynman integrals\\[2mm]
   {\em Nature of the physical problem\/}: 
	 Systematic expansion of higher transcendental functions in 
         a small parameter. 
         These expansions occur for example in the calculation of loop integrals
         in quantum field theory within dimensional regularization.\\[2mm]
   {\em Method of solution\/}: 
	 Algebraic manipulations of nested sums.\\[2mm] 
   {\em Restrictions on complexity of the problem\/}: 
	 Usually limited only by the available memory. \\[2mm]
   {\em Typical running time\/}:
         Depending on the complexitiy of the problem, see also sect. \ref{sec:checks}
         and \ref{sec:examples}.
\end{sloppypar}

\newpage

\reversemarginpar

{\bf\large LONG WRITE-UP}

\section{Introduction}
\label{sec:intro}

Higher transcendental function, like hypergeometric functions or Appell functions,
occur frequently in the calculation of Feynman integrals in quantum field theory.
Usually they appear with a small parameter $\eps$ in some of their arguments.
For example, within dimensional regularization 
\cite{'tHooft:1972fi}, 
the small parameter $\eps$ 
is given by the deviation
of $D$-dimensional space-time from the four-dimensional value: $D = 4 - 2 \eps$.
However, a solution in the form of transcendental functions is not yet useful, 
since possible poles in $\eps$ have not been made
explicit.
What is needed is a Laurent expansion in $\eps$.
In some cases this can be achieved with the help of an
integral representation \cite{Anastasiou:1999ui}.
However, this method is rather tedious and limited to cases where
an integral representation is known.
Recently, algorithms have been developed for the systematic expansion
of certain classes of transcendental functions \cite{Moch:2001zr}.
These algorithms are based on an algebra of nested sums.
In this paper we report on the implementation of these algorithms
into a computer library.\\
\\
Although these algorithms allow in principle the expansion to any desired
order, they grow exponentially with the complexity of the problem.
Therefore any implementation has to face the fact, that at intermediate
steps one possibly deals with very large expressions.
Commercial computer algebra systems are in these circumstances 
not always as reliable as one would
like them to be.
Computer algebra systems like FORM \cite{Vermaseren:2000nd}
or GiNaC \cite{Bauer:2000cp}, which have been developed
within the high-energy physics community, seem to be more appropriate.
Here we report on an implementation based on the GiNaC library.\\
\\
GiNaC is a library written in C++, which allows the symbolic manipulation
of expressions within the programming language C++.
There are several advantages to this approach:
C++ is a standardized programming language and supports
object-oriented programming.
The compilation of the programs leads to efficiency in the performance.
Furthermore, many development tools like for the automatic
generation of documentation are available.
In addition the source code is freely available.
\\
\\
The program library ``nestedsums'', which we describe here, extends the
capabilities of the GiNaC library towards the expansion of
transcendental functions. This is a domain, which up to now is usually not
available within commercial computer algebra systems.
\\
\\
The paper is organized as follows:
In the next section the definitions of nested sums and their main properties
are briefly recalled.
Section \ref{sec:design} gives an introduction to the design of the program
and aims at readers who wish to understand the source code.
Section \ref{sec:howto} is of a more practical nature and describes
how to install and use the program. 
Section \ref{sec:checks} deals with checks that we have performed and addresses
issues like performance. 
Some simple examples are given in section \ref{sec:examples}.
Finally a summary is provided in section \ref{sec:conclusions}.

\section{Theoretical Background}
\label{sec:back}

In this section we shortly review the algorithms for the systematic
expansion of transcendental functions.
They are described in detail in \cite{Moch:2001zr}.
They are based on a particular form of nested sums, which we call $Z$-sums.
$Z$-sums are defined by
\bq 
  Z(n;m_1,...,m_k;x_1,...,x_k) & = & \sum\limits_{n\ge i_1>i_2>\ldots>i_k>0}
     \frac{x_1^{i_1}}{{i_1}^{m_1}}\ldots \frac{x_k^{i_k}}{{i_k}^{m_k}}
\eq
and form a Hopf algebra.
If the sums go to Infinity ($n=\infty$) the $Z$-sums are the multiple polylogarithms of Goncharov \cite{Goncharov}:
\bq
\label{multipolylog}
Z(\infty;m_1,...,m_k;x_1,...,x_k) & = & \mbox{Li}_{m_k,...,m_1}(x_k,...,x_1).
\eq
For $x_1=...=x_k=1$ the definition reduces to the Euler-Zagier sums \cite{Euler,Zagier}:
\bq
Z(n;m_1,...,m_k;1,...,1) & = & Z_{m_1,...,m_k}(n).
\eq
For $n=\infty$ and $x_1=...=x_k=1$ the sum is a multiple $\zeta$-value \cite{Borwein}:
\bq
Z(\infty;m_1,...,m_k;1,...,1) & = & \zeta(m_k,...,m_1).
\eq
The multiple polylogarithms of Goncharov contain as the notation already suggests as subsets 
the classical polylogarithms 
$
\mbox{Li}_n(x)
$ 
\cite{lewin:book},
as well as
Nielsen's generalized polylogarithms \cite{Nielsen}
\bq
S_{n,p}(x) & = & \mbox{Li}_{1,...,1,n+1}(\underbrace{1,...,1}_{p-1},x),
\eq
the harmonic polylogarithms of Remiddi and Vermaseren \cite{Remiddi:1999ew}
\bq
\label{harmpolylog}
H_{m_1,...,m_k}(x) & = & \mbox{Li}_{m_k,...,m_1}(\underbrace{1,...,1}_{k-1},x)
\eq
and the two-dimensional harmonic polylogarithms introduced recently by
Gehrmann and Remiddi \cite{Gehrmann:2000zt}.
Euler-Zagier sums occur in the expansion of Gamma functions:
\bq
\label{expgamma}
\lefteqn{
\Gamma(n+\eps) = \Gamma(1+\eps) \Gamma(n) } & & \nonumber \\
& & \times \left( 1 + \eps Z_1(n-1) + \eps^2 Z_{11}(n-1) + \eps^3 Z_{111}(n-1) + ... 
+ \eps^{n-1} Z_{11...1}(n-1) \right). \nonumber \\
\eq
The usefulness of the $Z$-sums lies in the fact, that they interpolate between
Goncharov's multiple polylogarithms and Euler-Zagier sums.
In addition, the interpolation is compatible with the algebra structure.\\
\\
In addition to $Z$-sums, it is sometimes useful to introduce as well $S$-sums.
$S$-sums are defined by
\bq
S(n;m_1,...,m_k;x_1,...,x_k)  & = & 
\sum\limits_{n\ge i_1 \ge i_2\ge \ldots\ge i_k \ge 1}
\frac{x_1^{i_1}}{{i_1}^{m_1}}\ldots \frac{x_k^{i_k}}{{i_k}^{m_k}}.
\eq
The $S$-sums are closely related to the $Z$-sums, the difference being the upper summation boundary
for the nested sums: $(i-1)$ for $Z$-sums, $i$ for $S$-sums.
It is advantageous to introduce both $Z$-sums and $S$-sums, since some 
properties are more naturally expressed in terms of
$Z$-sums while others are more naturally expressed in terms of $S$-sums.
One can easily convert from the 
notation with $Z$-sums to the notation
with $S$-sums and vice versa.\\
\\
Since the $Z$-sums form an algebra, any product of two $Z$-sums with the same upper
summation bound $n$ can always be reduced to a sum of single $Z$-sums.
For example:
\bq
Z_{11}(n) \cdot Z_{1}(n) & = & Z_{21}(n) + Z_{12}(n) + 3 \; Z_{111}(n).
\eq
The multiplication property is at the core of all our algorithms.
In \cite{Moch:2001zr} we showed that four generic types of sums
can always be reduced to single $Z$-sums.
These four types are:\\
Type A:
\bq
\label{algo_A}
     \sum\limits_{i=1}^n \frac{x^i}{(i+c)^m} 
       \frac{\Gamma(i+a_1+b_1\varepsilon)}{\Gamma(i+c_1+d_1\varepsilon)} ...
       \frac{\Gamma(i+a_k+b_k\varepsilon)}{\Gamma(i+c_k+d_k\varepsilon)}
       Z(i+o-1,m_1,...,m_l,x_1,...,x_l),
\eq
Type B:
\bq
\label{algo_B}
\lefteqn{
     \sum\limits_{i=1}^{n-1} 
       \frac{x^i}{(i+c)^m} 
       \frac{\Gamma(i+a_1+b_1\varepsilon)}{\Gamma(i+c_1+d_1\varepsilon)}
...
       \frac{\Gamma(i+a_k+b_k\varepsilon)}{\Gamma(i+c_k+d_k\varepsilon)} \, 
       Z(i+o-1,m_1,...,m_l,x_1,...,x_l) }
\nonumber \\
& &
       \times
       \frac{y^{n-i}}{(n-i+c')^{m'}} 
       \frac{\Gamma(n-i+a_1'+b_1'\varepsilon)}{\Gamma(n-i+c_1'+d_1'\varepsilon)}
...
       \frac{\Gamma(n-i+a_{k'}'+b_{k'}'\varepsilon)}{\Gamma(n-i+c_{k'}'+d_{k'}'\varepsilon)}
\hspace*{4.0cm}
\nonumber \\[1ex]
& & \qquad\qquad \times
       Z(n-i+o'-1,m_1',...,m_{l'}',x_1',...,x_{l'}'),
\eq
Type C:
\bq
\label{algo_C}
\lefteqn{
       \mbox{} - \sum\limits_{i=1}^n 
       \left( \begin{array}{c} n \\ i \\ \end{array} \right)
       \left( -1 \right)^i
       \frac{x^i}{(i+c)^m} 
       \frac{\Gamma(i+a_1+b_1\varepsilon)}{\Gamma(i+c_1+d_1\varepsilon)}
       ...
       \frac{\Gamma(i+a_k+b_k\varepsilon)}{\Gamma(i+c_k+d_k\varepsilon)}
} & & \nonumber \\
& & \times
       S(i+o,m_1,...,m_l,x_1,...,x_l),
\hspace*{6cm}
\eq
Type D:
\bq
\label{algo_D}
\lefteqn{
     \mbox{} - \sum\limits_{i=1}^{n-1}
       \left( \begin{array}{c} n \\ i \\ \end{array} \right)
       \left( -1 \right)^i
       \frac{x^i}{(i+c)^m} 
       \frac{\Gamma(i+a_1+b_1\varepsilon)}{\Gamma(i+c_1+d_1\varepsilon)}
       ...
       \frac{\Gamma(i+a_k+b_k\varepsilon)}{\Gamma(i+c_k+d_k\varepsilon)}
} & &
\nonumber \\
& &
       \times
       S(i+o,m_1,...,m_l,x_1,...,x_l)
\nonumber \\
& & \times
       \frac{y^{n-i}}{(n-i+c')^{m'}} 
       \frac{\Gamma(n-i+a_1'+b_1'\varepsilon)}{\Gamma(n-i+c_1'+d_1'\varepsilon)}
       ...
       \frac{\Gamma(n-i+a_{k'}'+b_{k'}'\varepsilon)}{\Gamma(n-i+c_{k'}'+d_{k'}'\varepsilon)}
\nonumber \\
& & \times
       S(n-i+o',m_1',...,m_{l'}',x_1',...,x_{l'}').
\eq
Here, all $a_j$, $a_j'$, $c_j$  and $c_j'$ are integers, $c$, $c'$, 
are nonnegative integers and $o $,
$o' $ are integers. For sums of type A the upper summation limit $n$ may extend to Infinity.
The program library implements the algorithms for these four generic types of sums.

\section{Design of the program}
\label{sec:design}

C++ allows for object-oriented programming. Wherever it was compatible with
the required efficiency, we tried to write the program according to this
paradigm.
In C++ objects are represented by classes together with methods, which operate on them.
All our classes are derived from the fundamental class \v/basic/, which GiNaC provides.
In the following we describe shortly the main classes of the program.
The complete documentation is inserted as comments in the source code and can
easily be extracted and converted to latex or html format with tools like ``doxygen''.

\subsection{Letters}
\label{sec:letters}

\begin{figure}
\begin{center}
\begin{picture}(100,190)(0,0)
\Text(50,180)[c]{basic}
\Text(50,130)[c]{basic letter}
\Text(50,80)[c]{letter}
\Text(50,30)[c]{unit letter}
\ArrowLine(50,140)(50,170)
\ArrowLine(50,90)(50,120)
\ArrowLine(50,40)(50,70)
\end{picture}
\caption{\label{inheritance_letter} The inheritance diagram for the class basic letter and its derived classes.}
\end{center}
\end{figure}
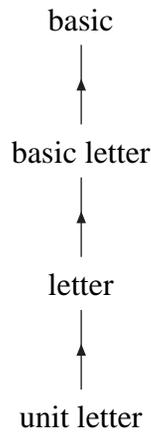

Expressions of the form 
\bq
\frac{x^i}{(i+c)^m}
\eq
occur quite often in the algorithms. They are represented by the specific class
\v/basic_letter/ which has as data members a scale $x$, a summation index $i$,
a degree $m$ and an offset $c$.
Two instances of this class with the same index $i$ 
can be multiplied, if they have the same offset or
if both degrees are integers, as for example in
\bq
\frac{x^i}{(i+c)^{m_1}} \frac{y^i}{(i+c)^{m_2}} 
 & = & \frac{(xy)^i}{(i+c)^{m_1+m_2}}, \nonumber \\
\frac{x^i}{(i+c_1)} \frac{y^i}{(i+c_2)} 
 & = & \frac{1}{(c_2-c_1)} \frac{x^i}{(i+c_1)}
       - \frac{1}{(c_2-c_1)} \frac{y^i}{(i+c_2)}.
\eq
In the second line the multiplication reduces to partial fractioning.
The multiplication operator is overloaded and multiplications are performed
whenever this is possible.
This is illustrated in the following code fragment:
\begin{verbatim}
  // construct two basic_letters
  ex l1 = basic_letter(x,m1,c,i);
  ex l2 = basic_letter(y,m2,c,i);

  // multiplication
  ex l = l1 * l2 ;
  // l is now basic_letter(x*y,m1+m2,c,i)
\end{verbatim}
If the offset $c$ equals zero, the object
\bq
\frac{x^i}{i^m}
\eq
is represented by the derived class \v/letter/. 
We will later see that $Z$-sums are constructed from these objects.
Note that \v/letter/ is derived from \v/basic_letter/ and
not vice versa. This is in accordance with the general rule, that a derived
class can be substituted everywhere, where the base class is required.
A further specialization occurs if the scale $x$ is equal to one.
The object
\bq
\frac{1}{i^m}
\eq
is represented by the class \v/unit_letter/, which is derived from
\v/letter/.
Fig. \ref{inheritance_letter} summarizes the inheritance relationships for letters.

\subsection{Nested sums}
\label{sec:sums}

\begin{figure}
\begin{center}
\begin{picture}(300,290)(0,0)
\Text(150,280)[c]{basic}
\Text(150,230)[c]{Z-sums}
\Text(50,180)[c]{multiple polylogs}
\Text(250,180)[c]{Euler-Zagier sums}
\Text(50,130)[c]{harmonic polylogs}
\Text(50,80)[c]{Nielsen polylogs}
\Text(50,30)[c]{classical polylogs}
\Text(250,130)[c]{multiple zeta values}
\ArrowLine(150,240)(150,270)
\ArrowLine(50,190)(130,220)
\ArrowLine(250,190)(170,220)
\ArrowLine(50,140)(50,170)
\ArrowLine(250,140)(250,170)
\ArrowLine(50,90)(50,120)
\ArrowLine(50,40)(50,70)
\ArrowLine(185,140)(100,170)
\end{picture}
\caption{\label{inheritance_Zsum} The inheritance diagram for $Z$-sums and derived classes.}
\end{center}
\end{figure}
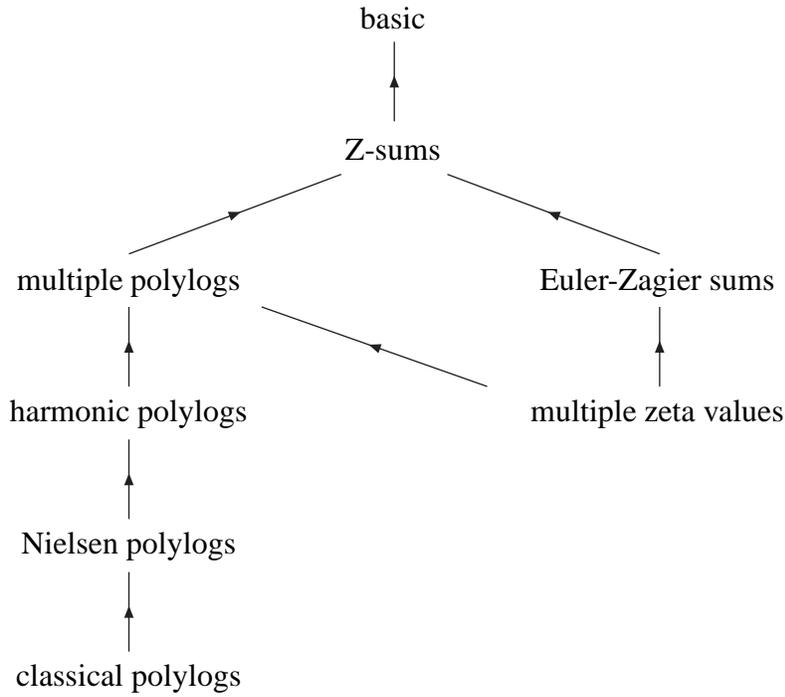

Probably the most important class within the program is the class \v/Zsum/ and its derived
classes. Fig. \ref{inheritance_Zsum} shows the inheritance relationships for the
derived classes, which correspond to the special cases discussed in eq. \ref{multipolylog}
to eq. \ref{harmpolylog}.
A \v/Zsum/ has the data members \v/n/
and \v/letter_list/. 
The former corresponds to the upper summation limit
whereas the later is a list of objects of the type \v/letter/.
The actual expressions for the summation indices occuring in the 
letters need not to be known and are therefore replaced by default
values.
The multiplication operator is overloaded and multiplications are immediately
performed, whenever this is possible.
The following code fragment illustrates this:
\begin{verbatim}
  ex l1 = unit_letter((ex) 1);

  ex Z1 = Zsum(n,lst(l1));     // Z1 is Z_{1}(n)
  ex Z11 = Zsum(n,lst(l1,l1)); // Z11 is Z_{11}(n)

  // multiplication
  ex res = Z1 * Z11;
  // res is now Z_{21}(n) + Z_{12}(n)+ 3*Z_{111}(n)
\end{verbatim}
A similar implementation for $S$-sums is provided by the class \v/Ssum/.

\subsection{Gamma functions}
\label{sec:gammas}

In the expansion of transcendental functions, Gamma functions occur always
in ratios.
The class \v/ratio_of_tgamma/ represents therefore the expression
\bq
\label{rtg}
\frac{\Gamma(1+b_2 \eps)}{\Gamma(1+b_1 \eps)}
\frac{\Gamma(i + a_1 + b_1 \eps)}{(i+a_2 + b_2 \eps)}.
\eq
The prefactor $\Gamma(1+b_2 \eps)/\Gamma(1+b_1 \eps)$ avoids an unnecessary proliferation
of Euler constants when this expression is expanded in $\eps$.
Apart from the obvious data members \v/a1/, \v/b1/, \v/a2/, \v/b2/, \v/index/ and
\v/expansion_parameter/ this class has two integer data members, \v/order/ and
\v/flag_expand_status/.
The \v/order/-variable specifies to which order the object shall eventually be
expanded.
The class is a ``smart'' class and expands itself into Euler-Zagier sums only if
\v/flag_expand_status/ is set.
The expansion is performed according to eq. \ref{expgamma}.\\
In addition there is a class \v/list_of_tgamma/, which is a container for
objects of the type \v/ratio_of_tgamma/ and represents a product of terms of the
form as in eq. \ref{rtg}.

\subsection{Algorithms}
\label{sec:algos}

The classes \v/transcendental_sum_type_A/ to \v/transcendental_sum_type_D/ 
implement the algorithms for the expansion of the sums given in eq. \ref{algo_A}
to eq. \ref{algo_D}.
These classes do the hard part of the work and are kept quite general.
Since they are kept quite general, they are not necessarily particular
user-friendly.
It is assumed that a user customizes these classes to his own needs and writes
his own interface.
An example is provided by the interface classes in the next subsection.
We shortly discuss the class \v/transcendental_sum_type_A/. The other
three classes have a similar structure.
The class \v/transcendental_sum_type_A/ represents the expression
\bq
\lefteqn{
    \frac{\Gamma(1+d_1\varepsilon)}{\Gamma(1+b_1\varepsilon)}
    ...
    \frac{\Gamma(1+d_k\varepsilon)}{\Gamma(1+b_k\varepsilon)}
} & & 
\nonumber \\
& & \times
     \sum\limits_{i=1}^n \frac{x^i}{(i+c)^m} 
       \frac{\Gamma(i+a_1+b_1\varepsilon)}{\Gamma(i+c_1+d_1\varepsilon)}
       ...
       \frac{\Gamma(i+a_k+b_k\varepsilon)}{\Gamma(i+c_k+d_k\varepsilon)}
       Z(i+o-1,m_1,...,m_l,x_1,...,x_l).
\eq
Apart from the prefactor 
$\Gamma(1+d_1\varepsilon)/\Gamma(1+b_1\varepsilon)...\Gamma(1+d_k\varepsilon)/\Gamma(1+b_k\varepsilon)$ this class corresponds exactly to eq. \ref{algo_A}.
The prefactor avoids that unnecessary terms involving Euler constants appear in the
expansion.
The data members of this class are: \v/n/ and \v/index/, representing the
upper summation limit and the summation index; \v/letter/, \v/lst_of_gammas/ and
\v/subsum/, representing the expression $x^i/(i+c)^m$, the Gamma functions and
the subsum, respectively. These members are of type
\v/basic_letter/, \v/list_of_tgamma/ and \v/Zsum/, respectively.
In addition there are the data members \v/expansion_parameter/, \v/order/ and
\v/flag_expand_status/.

\subsection{Interface classes}
\label{sec:interface}

The classes \v/transcendental_fct_type_A/ to \v/transcendental_fct_type_D/ 
provide a specific interface to the classes 
discussed above.
They represent:\\
Type A:
\bq
     \frac{\Gamma(d_1) ... \Gamma(d_n)}{\Gamma(d_1') ... \Gamma(d_{n'}')} 
      \sum\limits_{i=0}^\infty \frac{\Gamma(i+a_1) ... \Gamma(i+a_k)}{\Gamma(i+a_1') ... \Gamma(i+a_{k-1}')} \frac{x^i}{i!},
\eq
Type B:
\bq
\lefteqn{
     \frac{\Gamma(d_1) ... \Gamma(d_n)}{\Gamma(d_1') ... \Gamma(d_{n'}')} 
      \sum\limits_{i=0}^\infty \sum\limits_{j=0}^\infty 
       \frac{\Gamma(i+a_1) ... \Gamma(i+a_k)}{\Gamma(i+a_1') ... \Gamma(i+a_{k-1}')} 
       \frac{\Gamma(j+b_1) ... \Gamma(j+b_l)}{\Gamma(j+b_1') ... \Gamma(j+b_{l-1}')} 
} & &
\nonumber \\
& & \times
       \frac{\Gamma(i+j+c_1) ... \Gamma(i+j+c_m)}{\Gamma(i+j+c_1') ... \Gamma(i+j+c_{m}')} 
      \frac{x_1^i}{i!} \frac{x_2^j}{j!},
\eq
Type C:
\bq
     \frac{\Gamma(d_1) ... \Gamma(d_n)}{\Gamma(d_1') ... \Gamma(d_{n'}')} 
      \sum\limits_{i=0}^\infty \sum\limits_{j=0}^\infty 
       \frac{\Gamma(i+a_1) ... \Gamma(i+a_k)}{\Gamma(i+a_1') ... \Gamma(i+a_{k}')} 
       \frac{\Gamma(i+j+c_1) ... \Gamma(i+j+c_m)}{\Gamma(i+j+c_1') ... \Gamma(i+j+c_{m-1}')} 
      \frac{x_1^i}{i!} \frac{x_2^j}{j!},
\eq
Type D:
\bq
\lefteqn{
     \frac{\Gamma(d_1) ... \Gamma(d_n)}{\Gamma(d_1') ... \Gamma(d_{n'}')} 
      \sum\limits_{i=0}^\infty \sum\limits_{j=0}^\infty 
       \frac{\Gamma(i+a_1) ... \Gamma(i+a_k)}{\Gamma(i+a_1') ... \Gamma(i+a_{k}')} 
       \frac{\Gamma(j+b_1) ... \Gamma(j+b_l)}{\Gamma(j+b_1') ... \Gamma(j+b_{l}')} 
} & &
\nonumber \\
& & \times
       \frac{\Gamma(i+j+c_1) ... \Gamma(i+j+c_m)}{\Gamma(i+j+c_1') ... \Gamma(i+j+c_{m-1}')} 
      \frac{x_1^i}{i!} \frac{x_2^j}{j!}
\eq
and are modelled on generalizations of hypergeometric functions, the first
Appell function, the first Kamp\'e de F\'eriet function and the second
Appell function, respectively.
Note that the sums start at zero and not at one.
As an example we discuss the class \v/transcendental_fct_type_A/.
The constructor for type A is of the form
\begin{verbatim}
      transcendental_fct_type_A(x, a_num, a_denom, d_num, d_denom);
\end{verbatim}   
where \v/a_num/ is a list containing $a_1$ to $a_k$. For example for $k=4$ one has:
\begin{verbatim}
      ex a_num = lst(a1,a2,a3,a4);
\end{verbatim}   
Each $a_j$ has to be of the form $a_j = k+s \eps$, where $k$ is an integer and
$\eps$ the expansion parameter.
\v/a_denom/, \v/d_num/ and \v/d_denom/ are defined similar.
This class has a method
\begin{verbatim}
      ex set_expansion(const ex & eps, int order);
\end{verbatim}   
which expands the object to the desired order in $\eps$.
In addition there is a constructor, which expands the object directly.
The classes for type B to type D are implemented in complete analogy.

\section{How to Use the Library}
\label{sec:howto}

In this section we give indications how to install and use the program library.
Compilation of the package will build a (shared) library. The user can then write his own
programs, using the functions provided by the library by linking his executables
against the library.

\subsection{Installation}

The program library can be obtained from\\
\\
{\tt \hspace*{12pt} http://www.fis.unipr.it/\~{}stefanw/nestedsums}\\
\\
It requires the GiNaC library version 0.8.3 to be installed.
After unpacking, the library for nested sums is build by issuing
the commands
\begin{verbatim}
  ./configure 
  make
  make install
\end{verbatim}
There are various options which can be passed to the configure script,
an overview can be obtained with {\tt ./configure -{}-help}.\\
\\
After installation, the shell script \v/nestedsums-config/ can be used
to determine the compiler and linker command line options required
to compile and link a program with the nestedsums library.
For example, \v/nestedsums-config --cppflags/ will give the path to the header files
of the library, whereas \v/nestedsums-config --libs/ prints out the flags
necesarry to link a program against the library.

\subsection{Writing programs using the library}

Once the library is build and installed, it is ready to be used. Here 
is a small example program, which defines two Euler-Zagier sums,
multiplies them and prints out the result:
\begin{verbatim}
#include <iostream>
#include "ginac/ginac.h"
#include "nestedsums/nestedsums.h"

int main()
{
  using namespace GiNaC;

  symbol n("n");

  ex l1 = unit_letter((ex) 1);

  ex Z1 = Zsum(n,lst(l1));     
  ex Z11 = Zsum(n,lst(l1,l1)); 

  // multiplication
  ex res = Z1 * Z11;

  cout << res << endl;
}
\end{verbatim}
After compilation and linking against the GiNaC library and the nestedsums library,
one obtains an executable, which will print out
\begin{verbatim}
3*EZ(n,1,1,1)+EZ(n,2,1)+EZ(n,1,2)
\end{verbatim}
Here, \v/EZ/ is the output format for Euler-Zagier sums.

\subsection{Documentation}

The complete documentation of the program is inserted as comment lines in
the source code.
The documentation can be extracted from the sources
with the help of the documentation system ``doxygen'' \cite{doxygen}.
The program ``doxygen'' is freely available.
Issuing in the top-level build directory for the nestedsums library the commands
\begin{verbatim}
  doxygen Doxyfile
\end{verbatim}
will create a directory ``reference'' with the documentation in html and latex format.

\subsection{Interactive use}

For small problems it is desirable to avoid the editing/compilation cycle 
and to use the program interactively. In addition one would like the output
to be in high-quality fonts, like for example TeX fonts.
gTybalt \cite{gtybalt} is a program which allows one to do just this.
It is based on the C++ interpreter CINT \cite{cint} and uses the TeXmacs
editor \cite{texmacs} to display high-quality mathematical typesetting.
It has build-in support for the nestedsums library. To enable this support
one has to configure gTybalt with the option
\begin{verbatim}
  ./configure --with-nestedsums
\end{verbatim}
gTybalt is freely available under the GNU public license from\\
\\
{\tt \hspace*{12pt} http://www.fis.unipr.it/\~{}stefanw/gtybalt.html}

\section{Checks and performance}
\label{sec:checks}

The library comes with a testsuite, which performs several tests on the
implementation of the algorithms.
Non-trivial tests are checks of known identities, like for example
\bq
\lefteqn{
\hspace*{-6cm}
\left[ 1 + \eps Z_{1}(n) + \eps^2 Z_{11}(n) + ... + \eps^k Z_{11...1}(n) \right]
\left[ 1 - \eps S_{1}(n) + \eps^2 S_{11}(n) - ... + (-1)^k \eps^k S_{11...1}(n) \right]
} & & \nonumber \\
& = &
 1 + O(\eps^{k+1}).
\eq
This relation provides a check on the implementation of the conversion
between $Z$- and $S$-sums and on the multiplication of nested sums.
A second class of checks is obtained for finite sums by writing out the sums
explicitly and to compare the result of a symbolic manipulation with the
result obtained from the explicit expression.
Furthermore, identities obtained from partial integration also yield useful relations
which can be used for checks.
In the class of hypergeometric functions, Appell function and Kamp\'e de F\'eriet functions
we have for example the relations:
\bq
\lefteqn{
(c-1) {}_2F_1(a,b-1,c-1,x) - (c-1) {}_2F_1(a,b,c-1,x) + ax\; {}_2F_1(a+1,b,c,x) = 0,
} & &
\nonumber \\
\lefteqn{
(c_1-1) {}_3F_2(a,b_1-1,b_2,c_1-1,c_2,x) - (c_1-1) {}_3F_2(a,b_1,b_2,c_1-1,c_2,x) 
} & &
\nonumber \\ & &
      + x a \frac{b_2}{c_2} {}_3F_2(a+1,b_1,b_2+1,c_1,c_2+1,x) = 0,
\nonumber \\
\lefteqn{
(c-1) F_1(a-1,b_1,b_2,c-1,x_1,x_2) - (c-1) F_1(a,b_1,b_2,c-1,x_1,x_2) 
} & &
\nonumber \\ & &
    + x_1 b_1 F_1(a,b_1+1,b_2,c,x_1,x_2) 
    + x_2 b_2 F_1(a,b_1,b_2+1,c,x_1,x_2) = 0,
\nonumber \\
\lefteqn{
(c-1) S_1(a_1-1,a_2,b_1,c-1,c_1,x_1,x_2) - (c-1) S_1(a_1,a_2,b_1,c-1,c_1,x_1,x_2) 
} & &
\nonumber \\ & &
    + x_1 a_2 \frac{b_1}{c_1} S_1(a_1,a_2+1,b_1+1,c,c_1+1,x_1,x_2)
    + x_2 a_2 S_1(a_1,a_2+1,b_1,c,c_1,x_1,x_2) =  0,
\nonumber \\
\lefteqn{
(c_1-1) S_1(a_1,a_2,b_1-1,c,c_1-1,x_1,x_2) - (c_1-1) S_1(a_1,a_2,b_1,c,c_1-1,x_1,x_2) 
} & &
\nonumber \\ & &
     + x_1 a_1 \frac{a_2}{c} S_1(a_1+1,a_2+1,b_1,c+1,c_1,x_1,x_2) = 0,
\nonumber \\
\lefteqn{
(c_1-1) F_2(a,b_1-1,b_2,c_1-1,c_2,x_1,x_2) - (c_1-1) F_2(a,b_1,b_2,c_1-1,c_2,x_1,x_2) 
} & &
\nonumber \\ & &
     + x_1 a F_2(a+1,b_1,b_2,c_1,c_2,x_1,x_2) = 0.
\eq
For a given choice of the parameters, each term is expanded in $\eps$ independently
and one verifies that each relation yields zero up to the calculated order.\\
\\
In addition we considered the expression
\bq
I & = & 
 \frac{\Gamma(2m-2\eps-\nu_{1235})\Gamma(1+\nu_{1235}-2m+2\eps)
       \Gamma(2m-2\eps-\nu_{2345})\Gamma(1+\nu_{2345}-2m+2\eps)
      }{\Gamma(\nu_1) \Gamma(\nu_2) \Gamma(\nu_3) \Gamma(\nu_4) \Gamma(\nu_5)
        \Gamma(3m-3\eps-\nu_{12345})} \nonumber \\
& & 
 \times
 \frac{ \Gamma(m-\eps-\nu_5) \Gamma(m-\eps-\nu_{23})}{\Gamma(2m-2\eps-\nu_{235})}
 \left( -s_{123} \right)^{2m-2\eps-\nu_{12345}}
 \sum\limits_{i_1=0}^\infty
 \sum\limits_{i_2=0}^\infty
 \frac{x_1^{i_1}}{i_1!}
 \frac{x_2^{i_2}}{i_2!} \nonumber \\
& & \times \left[
   \frac{\Gamma(i_1+\nu_3) \Gamma(i_2+\nu_2) \Gamma(i_1+i_2-2m+2\eps+\nu_{12345})
                                             \Gamma(i_1+i_2-m+\eps+\nu_{235})
        }{\Gamma(i_1+1-2m+2\eps+\nu_{1235}) \Gamma(i_2+1-2m+2\eps+\nu_{2345})
          \Gamma(i_1+i_2+\nu_{23})}
 \right. \nonumber \\
 & &
 - x_1^{2m-2\eps-\nu_{1235}} 
  \nonumber \\
 & & \times
   \frac{\Gamma(i_1+2m-2\eps-\nu_{125}) \Gamma(i_2+\nu_2) \Gamma(i_1+i_2+\nu_4)
                                             \Gamma(i_1+i_2+m-\eps-\nu_1)
        }{\Gamma(i_1+1+2m-2\eps-\nu_{1235}) \Gamma(i_2+1-2m+2\eps+\nu_{2345})
          \Gamma(i_1+i_2+2m-2\eps-\nu_{15})}
 \nonumber \\
 & &
 - x_2^{2m-2\eps-\nu_{2345}}
 \nonumber \\
 & & \times
   \frac{\Gamma(i_1+\nu_3) \Gamma(i_2+2m-2\eps-\nu_{345}) \Gamma(i_1+i_2+\nu_1)
                                             \Gamma(i_1+i_2+m-\eps-\nu_4)
        }{\Gamma(i_1+1-2m+2\eps+\nu_{1235}) \Gamma(i_2+1+2m-2\eps-\nu_{2345})
          \Gamma(i_1+i_2+2m-2\eps-\nu_{45})}
 \nonumber \\
 & & 
 + x_1^{2m-2\eps-\nu_{1235}} x_2^{2m-2\eps-\nu_{2345}}
   \frac{\Gamma(i_1+2m-2\eps-\nu_{125}) \Gamma(i_2+2m-2\eps-\nu_{345}) 
        }{\Gamma(i_1+1+2m-2\eps-\nu_{1235}) \Gamma(i_2+1+2m-2\eps-\nu_{2345})
          } 
 \nonumber \\
 & & \left. \times
 \frac{
                                          \Gamma(i_1+i_2+2m-2\eps-\nu_{235})
                                          \Gamma(i_1+i_2+3m-3\eps-\nu_{12345})
      }{\Gamma(i_1+i_2+4m-4\eps-\nu_{12345}-\nu_5)}
 \right].
\eq
This expression corresponds to a two-loop integral occuring in the calculation
of the two-loop amplitude for $e^+ e^- \rightarrow \; \mbox{3 jets}$.
For the specific set of parameters $(\nu_1,\nu_2,...,\nu_5) = (1,1,1,1,1)$ and
$(1,1,1,1,2)$ it has been calculated by Gehrmann and Remiddi \cite{Gehrmann:2000zt} to
order $\eps^0$ and order $\eps^1$, respectively.
Our program allows us to obtain the result for arbitrary values for the $\nu_j$.
For the two specific cases we confirm the results of Gehrmann and Remiddi.\\
\\
Although the program has been written carefully and many bugs have been eliminated
during the debugging phase, there is no
guarantee that the program is bug-free.
It should be clear, that if the program is used to obtain results which are published
in scientific journals, it is still the responsibility of the user (and not of the author
of the program) to make sure that these results are correct.\\
\\
We would like to give some indications on the efficiency of our algorithms.
The inner parts of the algorithms are based on the multiplication of
nested sums and in this part efficiency is mandatory.
For the specific case of multiplication of harmonic sums there exists already
a program called ``summer'' which is written in FORM \cite{Vermaseren:1998uu}.
The comparison between the ``summer'' program and our program for the multiplication of two harmonic sums
\bq
S_{\underbrace{11...1}_{k}}(n) \cdot S_{\underbrace{11...1}_{k}}(n),
\eq
where each factor has $k$ ``one'''s in the index field is listed in table \ref{num_res1}.
\begin{table}
\label{num_res1}
\begin{center}
\begin{tabular}{|c|rrrrrrrrrr|} \hline
k    & 1 & 2 & 3 & 4 & 5 & 6 & 7 & 8 & 9 & 10 \\
\hline 
Form &  $< 1$ & $< 1$ & $< 1$ & $< 1$ & $< 1$ & 2 & 11 & 58 & 323 & 1816 \\ 
C++  &  $< 1$ & $< 1$ & $< 1$ & $< 1$ & $< 1$ & 1 &  5 & 30 & 180 & 1066 \\
\hline
\end{tabular}
\end{center}
\caption{
CPU time in seconds for the multiplication of two harmonic sums on
a 300 MHz Pentium II with 64 MB RAM.}
\end{table}
As can be seen from table \ref{num_res1}, our program is almost a factor two faster.
However, benchmarks tests should always be taken with a certain care.
Our program is fast, as long as all intermediate expressions fit into the available RAM.
No particular memory management facilities have been implemented.
If intermediate expressions exceed the available memory, the operating system
resorts to swapping pages, which slows down the execution time significantly.

\section{Examples}
\label{sec:examples}

In this section we give two example programs for the expansion of 
a hypergeometric function and a Appell function, together with some indications
on the CPU time required for the expansion up to a specific order.\\
\\
The following small program
\begin{verbatim}
#include <iostream>
#include "ginac/ginac.h"
#include "nestedsums/nestedsums.h"

int main()
{
  using namespace GiNaC;

  symbol a("a"), b("b"), c("c"), x("x"), eps("eps");

  int order = 4;

  // expands 2F1(a eps, b eps, 1 - c eps, x) in eps
  ex F21 = transcendental_fct_type_A(x,lst(a*eps,b*eps),lst(1-c*eps),
				     lst(1-c*eps),lst(a*eps,b*eps),
				     eps,order,expand_status::expansion_required);

  // some polishing
  F21 = convert_Zsums_to_standard_form(F21);

  cout << F21 << endl;
}
\end{verbatim}
expands the hypergeometric function
${}_2F_1(a\eps,b\eps,1-c\eps,x)$
in $\eps$ up to a given order. 
Running this program will print out
\begin{verbatim}
1+(eps^3*b*a*c+eps^3*b^2*a+eps^3*b*a^2)*S(1,2,x)+eps^3*b*a*Li(3,x)*c
+eps^2*b*a*Li(2,x)
\end{verbatim}
which agrees with the known expansion
\bq
{}_2F_{1}( a \eps, b \eps; 1 - c \eps; x) & = & 
  1 + a b \;\mbox{Li}_2(x) \eps^2
  + a b \left( c \;\mbox{Li}_3(x) + ( a+b+c ) \;\mbox{S}_{1,2}(x) \right) \eps^3
  + O(\eps^4). \nonumber \\
\eq
The function \v/convert_Zsums_to_standard_form/ brings an expression involving 
$Z$-sums into a standard form. It first removes all $Z$-sums with non-positive
indices. An example
would be 
\bq
\mbox{Li}_{0}(x) & = & \sum\limits_{i=1}^\infty x^i = \frac{x}{1-x}.
\eq
These sums are rather trivial and can be eliminated.
It further collects the coefficients of each $Z$-sum and brings the coefficients into a normal
form.
\begin{table}
\label{num_res2}
\begin{center}
\begin{tabular}{|c|rrrrrrrrrrr|} \hline
order    & 1 & 2 & 3 & 4 & 5 & 6 & 7 & 8 & 9 & 10 & 11 \\
\hline 
${}_2F_1( a \eps, b \eps; 1 - c \eps; x)$  & < 1 & < 1 & 1 & 2 & 4 & 8 &  16 & 31 & 59 & 121 & 285 \\
\hline
\end{tabular}
\end{center}
\caption{
CPU time in seconds for the expansion of a hypergeometric function
on a 300 MHz Pentium II with 64 MB RAM.}
\end{table}
Table \ref{num_res2} shows the dependency of the CPU time for the expansion
of the hypergeometric function on the chosen order.\\
\\
As a second example we give a short program, which expands the second 
Appell function $F_2(1,1,\eps;1+\eps,1-\eps;x,y)$ in $\eps$:
\begin{verbatim}
#include <iostream>
#include "ginac/ginac.h"
#include "nestedsums/nestedsums.h"

int main()
{
  using namespace GiNaC;

  symbol x("x"), y("y"), eps("eps");

  int order = 2;

  // expands F2(1,1,eps;1+eps,1-eps;x,y) in eps
  ex F2 = transcendental_fct_type_D(x,y,lst(1),lst(1+eps),lst(eps),lst(1-eps),
				    lst(1),lst(),lst(1+eps,1-eps),lst(1,1,eps),
				    eps,order,expand_status::expansion_required);

  // some polishing
  F2 = convert_Zsums_to_standard_form(F2);

  cout << F2 << endl;
}
\end{verbatim}
Running this program will print out
\begin{verbatim}
-(-1+x)^(-1)+2*Li(1,x)*(-1+x)^(-1)*eps-Li(1,y+x)*(-1+x)^(-1)*eps
\end{verbatim}
which corresponds to the expansion
\bq
F_2(1,1,\eps;1+\eps,1-\eps;x,y) & = & 
 \frac{1}{1-x}
 + \frac{1}{1-x} \left( 2 \ln(1-x) - \ln(1-x-y) \right) \eps 
 + O(\eps^2).
\nonumber \\
\eq
\begin{table}
\label{num_res3}
\begin{center}
\begin{tabular}{|c|rrrrrr|} \hline
order    & 1 & 2 & 3 & 4 & 5 & 6 \\
\hline 
$F_2(1,1,\eps;1+\eps,1-\eps;x,y)$  & < 1 & < 1 & 2 & 9 & 49 & 279 \\
\hline
\end{tabular}
\end{center}
\caption{CPU time in seconds for the expansion of the second Appell function
on a 300 MHz Pentium II with 64 MB RAM.}
\end{table}
Table \ref{num_res3} shows the dependency of the CPU time for the expansion
of the second Appell function on the chosen order.
The algorithm for the expansion of the Appell function is more complex
compared to the one for the hypergeometric function, a fact which is also
reflected in the necessary CPU time.

\section{Summary}
\label{sec:conclusions}

In this paper we have described the program library ``nestedsums''.
This library can be used for the symbolic expansion of a certain class
of transcendental functions and can be useful to scientists in all fields.
In particular these sort of expansions are required in the calculation
of higher order corrections to scattering processes in high energy physics.
The library is written in C++ and uses the GiNaC library.

\subsection*{Acknowledgements}
\label{sec:acknowledgements}

I would like to thank Thomas Gehrmann for providing me with the program tdhpl
\cite{Gehrmann:2001jv}.



\begin{thebibliography}{10}

\bibitem{'tHooft:1972fi}
G.~'t~Hooft and M.~J.~G. Veltman,
\newblock Nucl. Phys. {\bf B44}, 189 (1972);
\\
C.~G. Bollini and J.~J. Giambiagi,
\newblock Nuovo Cim. {\bf B12}, 20 (1972);
\\
G.~M. Cicuta and E.~Montaldi,
\newblock Nuovo Cim. Lett. {\bf 4}, 329 (1972).

\bibitem{Anastasiou:1999ui}
C.~Anastasiou, E.~W.~N. Glover, and C.~Oleari,
\newblock Nucl. Phys. {\bf B572}, 307 (2000), hep-ph/9907494.

\bibitem{Moch:2001zr}
S.~Moch, P.~Uwer, and S.~Weinzierl,
\newblock (2001), hep-ph/0110083.

\bibitem{Vermaseren:2000nd}
J.~A.~M. Vermaseren,
\newblock (2000), math-ph/0010025.

\bibitem{Bauer:2000cp}
C.~Bauer, A.~Frink, and R.~Kreckel,
\newblock J. Symbolic Computation {\bf 33}, 1 (2002), cs.sc/0004015;
\\
the GiNaC library is available at http://www.ginac.de.

\bibitem{Goncharov}
A.~B. Goncharov,
\newblock Math. Res. Lett. {\bf 5}, 497 (1998), available at\\
http://www.math.uiuc.edu/K-theory/0297.

\bibitem{Euler}
L.~Euler,
\newblock Novi Comm. Acad. Sci. Petropol. {\bf 20}, 140 (1775).

\bibitem{Zagier}
D.~Zagier,
\newblock First European Congress of Mathematics, Vol. II, Birkhauser, Boston ,
  497 (1994).

\bibitem{Borwein}
J.~M.~Borwein, D. M.~Bradley, D.~J.~Broadhurst and P.~Lisonek,
\newblock math.CA/9910045.

\bibitem{lewin:book}
L.~Lewin,
{\it Polylogarithms and associated functions},
\newblock  (North Holland, Amsterdam, 1981).

\bibitem{Nielsen}
N.~Nielsen,
\newblock Nova Acta Leopoldina (Halle) {\bf 90}, 123 (1909).

\bibitem{Remiddi:1999ew}
E.~Remiddi and J.~A.~M. Vermaseren,
\newblock Int. J. Mod. Phys. {\bf A15}, 725 (2000), hep-ph/9905237.

\bibitem{Gehrmann:2000zt}
T.~Gehrmann and E.~Remiddi,
\newblock Nucl. Phys. {\bf B601}, 248 (2001), hep-ph/0008287.

\bibitem{doxygen}
D.~van Heersch,
\newblock http://www.stack.nl/{\~{}}dimitri/doxygen.

\bibitem{gtybalt}
S.~Weinzierl and R.~Marani,
\newblock http://www.fis.unipr.it/{\~{}}stefanw/gtybalt.html.

\bibitem{cint}
M.~Goto,
\newblock http://root.cern.ch/root/Cint.html.

\bibitem{texmacs}
J.~van~der Hoeven,
\newblock http://www.texmacs.org.

\bibitem{Vermaseren:1998uu}
J.~A.~M. Vermaseren,
\newblock Int. J. Mod. Phys. {\bf A14}, 2037 (1999), hep-ph/9806280.

\bibitem{Gehrmann:2001jv}
T.~Gehrmann and E.~Remiddi,
\newblock (2001), hep-ph/0111255.



\end{thebibliography}
\end{document}